\documentclass[aps,prd,nofootinbib,amsmath,amssymb,superscriptaddress,showpacs,showkeys,twocolumn,10pt]{revtex4}%superscriptaddress,showpacs,
\usepackage{graphicx}
\usepackage{dcolumn}
\usepackage{bm}
\usepackage{amssymb}
\usepackage{latexsym}
\usepackage{booktabs}
\usepackage{amsmath}
\usepackage{multirow}
\usepackage[colorlinks=true, linkcolor=red, citecolor=blue]{hyperref}

\newcommand{\be}{\begin{equation}}
\newcommand{\ee}{\end{equation}}
\newcommand{\bq}{\begin{eqnarray}}
\newcommand{\eq}{\end{eqnarray}}

\begin{document}

\title{Probing the sign-changeable interaction between dark energy and dark matter with current observations}

\author{Juan-Juan Guo}
\affiliation{Department of Physics, College of Sciences, Northeastern University, Shenyang
110004, China}
\author{Jing-Fei Zhang}
\affiliation{Department of Physics, College of Sciences, Northeastern University, Shenyang
110004, China}
\author{Yun-He Li}
\affiliation{Department of Physics, College of Sciences, Northeastern University, Shenyang
110004, China}
\author{Dong-Ze He}
\affiliation{Department of Physics, College of Sciences, Northeastern University, Shenyang 110004, China}
\author{Xin Zhang\footnote{Corresponding author.}}\email{zhangxin@mail.neu.edu.cn}
\affiliation{Department of Physics, College of Sciences,
Northeastern University, Shenyang 110004, China}
\affiliation{Center for High Energy Physics, Peking University, Beijing 100080, China}

\begin{abstract}

We consider the models of vacuum energy interacting with cold dark matter in this study, in which the coupling can change sigh during the cosmological evolution. We parameterize the running coupling $b$ by the form $b(a)=b_0a+b_e(1-a)$, where at the early-time the coupling is given by a constant $b_{e}$ and today the coupling is described by another constant $b_{0}$. We explore six specific models with (i) $Q=b(a)H_0\rho_0$, (ii) $Q=b(a)H_0\rho_{\rm de}$, (iii) $Q=b(a)H_0\rho_{\rm c}$, (iv) $Q=b(a)H\rho_0$, (v) $Q=b(a)H\rho_{\rm de}$, and (vi) $Q=b(a)H\rho_{\rm c}$. The current observational data sets we use to constrain the models include the JLA compilation of type Ia supernova data, the Planck 2015 distance priors data of cosmic microwave background observation, the baryon acoustic oscillations measurements, and the Hubble constant direct measurement. We find that, for all the models, we have $b_0<0$ and $b_e>0$ at around the 1$\sigma$ level, and $b_0$ and $b_e$ are in extremely strong anti-correlation. Our results show that the coupling changes sign during the evolution at about the 1$\sigma$ level, i.e., the energy transfer is from dark matter to dark energy when dark matter dominates the universe and the energy transfer is from dark energy to dark matter when dark energy dominates the universe.

\end{abstract}


\maketitle

\section{Introduction}
\label{sec1}

Through great efforts of several decades by the whole community of cosmology, the basic framework of a standard cosmological model has been established, for which the most important feature is that the model involves dark energy, dark matter, and inflation, but by far we still do not clearly know the physical nature of dark energy and dark matter, as well as the physical details of inflationary process. A prototype of the standard cosmological model is the so-called $\Lambda$ cold dark matter ($\Lambda$CDM) model, in which dark energy is described by a cosmological constant $\Lambda$, dark matter is cold (i.e., non-relativistic particles), and inflation yields  adiabatic, Gaussian, nearly scale-invariant primordial density perturbations and possibly detectable primordial gravitational waves. The Planck satellite mission released the unprecedentedly accurate data of cosmic microwave background (CMB) temperature and polarization power spectra, in combination with other astrophysical observations, strongly favoring a 6-parameter base $\Lambda$CDM model \cite{Ade:2015xua}. But, it is hard to believe that the base $\Lambda$CDM model with only 6 primary parameters is the final model of cosmology. It is necessary to extend the base $\Lambda$CDM model in several aspects, but the current observational data are not sufficiently accurate to provide compelling evidence for these possible extensions.

For dark energy, though the cosmological constant $\Lambda$ can provide a simple, nice explanation, it always suffers from the severe theoretical challenges, such as the fine-tuning and coincidence problems \cite{Weinberg:1988cp,Sahni:1999gb,
Peebles:2002gy,Bean:2005ru,Copeland:2006wr,
Sahni:2006pa,Kamionkowski:2007wv,
Frieman:2008sn,Li:2011sd}. To extend the $\Lambda$CDM cosmology on this aspect, there are two possibilities, i.e., dynamical dark energy and modified gravity  (MG) theories (see Ref.~\cite{Weinberg:2012es} for a recent review). If the Einstein's general relativity (GR) is valid on all the scales of the universe, then $\Lambda$ can be extended to some dynamical dark energy with the equation-of-state parameter (EoS) $w$ not exactly equal to $-1$ but dynamically evolutionary. A typical example of this kind is provided by a spatially homogenous, slowly-rolling scalar field, called quintessence. A dynamical dark energy, compared to the cosmological constant, can yield a different expansion history of the universe, but yield a similar growth history of structure. On the other hand, the possibility of GR breaking down on the cosmological scales exists, and actually some models of MG can indeed produce an ``effective dark energy'' to explain the accelerated expansion of the universe. In general, the MG models can yield a similar expansion history but yield a quite different growth history, compared to the $\Lambda$CDM cosmology, in particular, the predicted linear growth of structure in the MG models is usually scale-dependent, fairly different from the prediction in GR. In some sense, the MG theory can be viewed as a model of some scalar field interacting with other components including dark matter, baryons, neutrinos, and photons, through mediating some scalar degrees of freedom, also known as ``the fifth force''. Differentiating the cases of dark energy and MG becomes nowadays one of the most important missions in the research area of cosmic acceleration.

However, it must be pointed out that actually there is also another important theoretical possibility that dark energy directly interacts with dark matter \cite{Amendola:1999er,Amendola:2001rc,Comelli:2003cv,Cai:2004dk,Zhang:2005rg,Zimdahl:2005bk,Wang:2006qw,Guo:2007zk,Bertolami:2007zm,Boehmer:2008av,Zhang:2007uh,He:2008tn,He:2009mz,He:2009pd,Koyama:2009gd,Li:2009zs,Xia:2009zzb,Zhang:2009qa,Li:2010ak,He:2010im,Chen:2011rz,Clemson:2011an,Wang:2013qy,Fu:2011ab,Wang:2014oga,Geng:2015ara,Yin:2015pqa,Feng:2016djj,Fan:2015rha,Murgia:2016ccp,Sola:2016jky,Sola:2016ecz,Sola:2016zeg,Sola:2017jbl,Pourtsidou:2016ico,Costa:2016tpb,Xia:2016vnp,vandeBruck:2016hpz,Kumar:2016zpg,Kumar:2017dnp,Santos:2017bqm,Valiviita:2008iv,Li:2014eha,Li:2014cee,Li:2015vla,Guo:2017hea,Zhang:2017ize,Zhang:2012uu,Zhang:2013lea,Li:2013bya,Zhang:2004gc,Cai:2009ht,Li:2011ga} (see also Ref.~\cite{Wang:2016lxa} for a recent review). That is to say, perhaps the fifth force only exists between dark energy and dark matter, and the normal baryons and radiation cannot perceive the fifth force. This possibility is tantalizing, not only because the interacting dark energy models can provide a scheme for solving (or alleviating) the cosmic coincidence problem theoretically through the attractor solution, but also because the possibility of dark energy interacting with dark matter itself indeed needs rigorous observational tests. In the MG models, the baryon matter can also perceive the fifth force, which leads to strong constraints from small scales such as the solar system, and thus the screening (or chameleon) mechanism is necessary for this kind of models. In particular, in the MG models, the linear matter perturbations are dependent on scales or environments. But in the interacting dark energy models, since the baryon matter does not perceive the fifth force, the screening mechanism is not needed.

Using the observational data to test the interacting dark energy scenario is an important mission. Difficulties in this area lie in the following aspects: (\romannumeral1) Owing to the lack of the theoretical knowledge of microscopic origin of such an interaction, one has to first phenomenologically propose a specific interacting dark energy model and then tests its theoretical and observational consequences. Thus, the studies are more or less model-dependent. (\romannumeral2) When investigating the matter perturbations in the interacting dark energy scenario, the problem of perturbation divergence on super-horizon scales at early times was found \cite{Valiviita:2008iv}. Fortunately, a parameterized post-Friedmann framework for interacting dark energy \cite{Li:2014eha,Li:2014cee,Li:2015vla,Guo:2017hea,Zhang:2017ize} has been established, which can be used to successfully solve the problem of perturbation divergence in the interacting dark energy scenario. But this issue shows that we are indeed ignorant of the nature of dark energy, in particular, we do not know at all how the acoustic waves propagate in the dark energy fluid. (\romannumeral3) The observational features are not very distinct for the interacting dark energy scenario. The interaction between dark energy and dark matter directly changes the expansion history of the universe (due to the combination of the Friedmann equation with the modified continuity equations of dark energy and dark matter), and thus the growth of structure is also changed, but the change is not remarkable. The reason is that the change of growth originates from the change of expansion history in the interacting dark energy scenario, which leads to the fact that the growth of structure in this scenario is still scale-independent (for the linear regime). This is different from the MG theory in which the growth of structure is scale-dependent.

Under such circumstances, probing the interaction between dark energy and dark matter with observations is very difficult, but it is fairly meaningful. To investigate the interacting dark energy scenario, one starts from the continuity equations for dark sectors by assuming an energy transfer between them,
\begin{equation}\label{de}
\dot{\rho}_{\rm de}+3H(1+w)\rho_{\rm de}=Q,
\end{equation}
\begin{equation}\label{cdm}
\dot{\rho}_{\rm c}+3H\rho_{\rm c}=-Q,
\end{equation}
where $\rho_{\rm de}$ and $\rho_{\rm c}$ are the energy densities of dark energy and cold dark matter, respectively, $H=\dot{a}/a$ is the Hubble parameter, with $a$ the scale factor of the universe and the dot the derivative with respect to the cosmic time $t$, $w=p_{\rm de}/\rho_{\rm de}$ is the EoS of dark energy, and $Q$ is the energy (density) transfer rate between dark sectors. By our convention, $Q>0$ means that the energy transfer is from dark matter to dark energy, and $Q<0$ means that the energy transfer is from dark energy to dark matter. This convention is in accordance with our recent a series of works on the interacting dark energy scenario \cite{Li:2014eha,Li:2014cee,Li:2015vla,Guo:2017hea,Zhang:2017ize}. Since we do not theoretically understand the microscopic origin of such an interaction, we have to study this issue from a phenomenological perspective. Usually, the phenomenological form of $Q$ is designed to be proportional to the density of dark sectors, i.e., $Q\propto \rho$, where the assumptions of $\rho=\rho_{\rm de}$, $\rho=\rho_{\rm c}$, and $\rho=\rho_{\rm de}+\rho_{\rm c}$ are most widely accepted. Such a design is made by consulting the models of reheating, of dark matter decaying into radiation, and of curvaton decay, in which the energy density transfer rate (namely the interaction term $Q$) has the form proportional to the density of one species. Of course, there are also some more exotic, complex forms, such as $Q\propto \rho_{\rm de}^\alpha \rho_{\rm c}^\beta$ \cite{Zhang:2012uu,Zhang:2013lea} or $Q\propto \rho_{\rm de} \rho_{\rm c}/(\rho_{\rm de}+\rho_{\rm c})$ \cite{Li:2013bya,Zhang:2004gc}, etc. One often considers two classes of models, i.e., $Q=bH\rho$ and $Q=bH_0\rho$, where $b$ is a dimensionless coupling parameter. The former includes a Hubble parameter, which is mathematically simpler for related calculations and thus much more popular in this area, and the latter excludes the Hubble parameter in its expression (the inclusion of the Hubble constant $H_0$ is for a dimensional reason), for which an argument is that the local interaction event should not be relevant to the global expansion of the universe. Both classes of forms have been extensively studied.

In fact, one can be aware that in the research area of interacting dark energy the models are significantly dependent on the man-made choice of the special interaction forms. Namely, the theoretical predictions and observational consequences of such a scenario are model-dependent. Actually, an important possibility that the interaction changes sign (i.e., the energy transfer direction changes) during the cosmological evolution exists; but the above-mentioned phenomenological models cannot describe this situation.

In Ref.~\cite{Cai:2009ht}, Cai and Su first investigated the case of sign-changeable interaction. They did not choose a special phenomenological form of interaction, but adopted a piecewise scheme in which the whole redshift range is divided into a few bins and the coupling $\delta(z)$ (in their model $Q=3H\delta$) is set to be a constant in each bin. They found that $\delta(z)$ is likely to cross the noninteracting line ($\delta=0$) according to the result of fitting to the observational data at that time. But the limitation of the work is that the sign-change behavior is based on the best-fit $\delta(z)$ and thus is not conclusive, because the errors of the fit results weaken the conclusion to a great degree. For a piecewise parametrization approach, the observational data in the current stage are hard to determine more than two parameters. For this reason, two authors of this paper (Li and Zhang) \cite{Li:2011ga} subsequently investigated this issue in a different approach. See Refs.~\cite{Wei:2010fz,Wei:2010cs,Zhang:2012sya,Sun:2010vz,Forte:2013fua,xu13a,xu13b,huang2014,Zhang:2014zfa,Xu:2015ata,Xi:2015qua,Zadeh:2016vgc,Zadeh:2017zke} for other works on the models of sign-changeable interaction.

In Ref.~\cite{Li:2011ga}, Li and Zhang proposed a parametrization form for the interaction term, $Q(a)=3b(a)H_0\rho_0$,
where $b(a)$ is a dimensionless coupling that is variable during the cosmological evolution ($a$ is the scale factor of the universe), $H_0$ is the Hubble constant, and $\rho_0=\rho_{\rm de0}+\rho_{\rm c0}$ is the present-day density of dark sectors. It should be pointed out that here the occurrence of $H_0$ and $\rho_0$ is only for a dimensional reason. Since the evolution of the interaction term $Q$ is completely described by the running of the coupling $b$, this is called the ``running coupling'' scenario in Ref.~\cite{Li:2011ga}. Furthermore, Li and Zhang \cite{Li:2011ga} proposed a parametrization form for the coupling,
\begin{equation}\label{b}
b(a)=b_0a+b_e(1-a),
\end{equation}
where $b_0$ and $b_e$ are two dimensionless parameters. It can be seen that the coupling $b$ is described by $b_0$ at the late times and determined by $b_e$ at the early times. So the entire evolution of $b(a)$ is totally characterized by the two parameters, $b_0$ and $b_e$, and Eq.~(\ref{b}) continuously connects the early-time and late-time behaviors. It was shown in Ref.~\cite{Li:2011ga} that the observational data at that time, including the CMB data from WMAP 7-year observation, the type Ia supernova (SNIa) data from Union2, the baryon acoustic oscillation (BAO) data from SDSS DR7, the Hubble expansion rate data, and the X-ray gas mass fraction data from Chandra measurement, favor a time-varying vacuum scenario at about the 1$\sigma$ level, in which the coupling $b(z)$ crosses the non-interacting line ($b=0$) at around $z=0.2\sim 0.3$ during the cosmological evolution.

In this paper, we wish to further explore the sign-changeable interacting dark energy scenario by using the latest observational data. In the current situation, the Planck CMB data as well as numerous more accurate astrophysical data were released, and thus we would like to examine whether or not the conclusion of Ref.~\cite{Li:2011ga} still holds. Moreover, in order to get more convincing and conclusive evidence, we wish to examine more forms of $Q$. Thus, in this work, we use the latest cosmological observations (including the Planck 2015 observation) to explore the sign-changeable models of interacting dark energy.

\section{Models}

In a spatially flat Friedmann-Roberston-Walker universe, the Friedmann equation reads

\begin{equation}
  3M_{\rm pl}^{2}H^{2}=\rho_{\rm c}+\rho_{\rm b}+\rho_{\rm r}+\rho_{\rm de},
\end{equation}
where $\rho_{\rm c}$, $\rho_{\rm b}$, $\rho_{\rm r}$, and $\rho_{\rm de}$ are the energy densities of cold dark matter, baryon, radiation, and dark energy, respectively. As usual, we introduce the fractional energy densities, $\Omega_{i}=\rho_{i}/3M_{\rm pl}^{2}H^{2}$, with $3M_{\rm pl}^{2}H^{2}$ defined as the critical density of the universe. So, we have
\begin{equation}
  \Omega_{\rm c}+\Omega_{\rm b}+\Omega_{\rm r}+\Omega_{\rm de}=1.
\end{equation}
If there exists some direct, non-gravitational interaction between dark energy (DE) and dark matter (DM), neither of them will evolve independently, and thus the continuity equations of them are given by Eqs.~(\ref{de}) and (\ref{cdm}). And, the energy conservation equations for the baryon and radiation components are usual, i.e., $\dot{\rho}_{\rm b}+3H\rho_{\rm b}=0$ and $\dot{\rho}_{\rm r}+4H\rho_{\rm r}=0$.

In this work, we consider the following six cases of the interaction term $Q$:
\begin{equation}\label{q1}
Q=b(a)H_{0}\rho_{\rm 0},
\end{equation}
\begin{equation}\label{q2}
Q=b(a)H_{0}\rho_{\rm de},
\end{equation}
\begin{equation}\label{q3}
Q=b(a)H_{0}\rho_{\rm c},
\end{equation}
\begin{equation}\label{q4}
Q=b(a)H\rho_{\rm 0},
\end{equation}
\begin{equation}\label{q5}
Q=b(a)H\rho_{\rm de},
\end{equation}
\begin{equation}\label{q6}
Q=b(a)H\rho_{\rm c}.
\end{equation}
Note that Eq.~(\ref{q1}) is just the one that considered in Ref.~\cite{Li:2011ga}. Here $\rho_{0}=\rho_{\rm c0}+\rho_{\rm de0}$ in Eqs.~(\ref{q1}) and (\ref{q4}), which is only for a convenient dimensional consideration. Eqs.~(\ref{q1})--(\ref{q3}) exclude the Hubble parameter and Eqs.~(\ref{q4})--(\ref{q6}) include the Hubble parameter in the expressions. In these models, the running coupling $b(a)$ is given by Eq.~(\ref{b}), i.e., $b(a)=b_0a+b_e(1-a)$.

In this work, for dark energy, we only consider the vacuum energy with $w=-1$. This is because we only wish to extend the base $\Lambda$CDM model in a minimal way. In this way, we can simply discuss the effects of interaction between DE and DM, but exclude the degeneracy with dynamical dark energy parameters. The models of vacuum energy interacting with cold dark matter investigated in this paper are abbreviated as ``I$\Lambda$CDM'' models. This does not mean that the vacuum energy density maintains as a constant in space and time; actually, in this case, the vacuum energy density $\rho_v$ is now a dynamical quantity because $\dot{\rho}_v=Q$ [see Eq.~(\ref{de}), with $w=-1$]. Now, the vacuum energy density is not truly equivalent to the cosmological constant $\Lambda$ any more, although $w=-1$ still holds. One should keep in mind that ``I$\Lambda$CDM'' is only an abbreviation.

For convenience, in the following, the I$\Lambda$CDM models with the interaction term $Q(a)$ described by Eqs.~(\ref{q1})--(\ref{q6}) are denoted as I$\Lambda$CDM1, I$\Lambda$CDM2, I$\Lambda$CDM3, I$\Lambda$CDM4, I$\Lambda$CDM5, and I$\Lambda$CDM6, respectively.

\section{Method and data}

In this work, we use a variety of cosmological data sets to constrain the phenomenological I$\Lambda$CDM models and obtain the best-fit parameters and likelihoods by performing a Markov-chain Monte Carlo (MCMC) analysis with the publicly available \texttt{CosmoMC} \cite{Lewis:2002ah} package.

We apply the $\chi^{2}$ statistic to fit the cosmological models to the observational data. The $\chi^{2}$ function is given by
\begin{equation}
  \chi_{\xi}^{2}=\frac{(\xi_{\rm th}-\xi_{\rm obs})^{2}}{\sigma_{\xi}^{2}},
\end{equation}
where $\xi_{\rm obs}$ is the experimentally measured value, $\xi_{\rm th}$ is the theoretically predicted value, and $\sigma_{\xi}$ is the standard deviation of the physical quantity $\xi$. The total $\chi^{2}$ is the sum of all $\chi_{\xi}^{2}$, and in this paper, we perform a joint SN+CMB+BAO+$H_{0}$ fit, thus the total $\chi^{2}$ can be given by
\begin{equation}
  \chi^{2}=\chi^{2}_{\rm SN}+\chi^{2}_{\rm CMB}+\chi^{2}_{\rm BAO}+\chi^{2}_{\rm H_{0}}.
\end{equation}

Next, we will compactly describe the observational data sets mentioned above. We employ the JLA SN data, the Planck CMB distance priors data, the BAO data and the $H_{0}$ direct measurement. It should be mentioned that, in our fit analysis, these various observations are well consistent with each other. We do not consider the cosmological perturbations in the DE models. Since the smooth dark energy affects the growth of structure only through the cosmic expansion history, different smooth dark energy models would result in almost the same growth history of structure. Even though the interaction is introduced, the effects on perturbations are small, relative to those on the expansion history. Thus, in this paper, we only consider the observational data of expansion history, i.e., those depicting the distance-redshift relations.

\emph{The SN data}: We use the JLA compilation of type Ia supernova, which is from a joint analysis of type Ia supernova observations in the redshift range of $z\in[0.001,1.30]$. It consists of 740 Ia supernovae, containing several low-redshift samples obtained from the SDSS and SNLS, and a few high-redshift samples from the HST \cite{Betoule:2014frx}.

\emph{The CMB data}: We use the ``Planck distance priors" from the Planck 2015 data \cite{Ade:2015rim}. Note here that dark energy could affect the CMB through not only the comoving angular diameter distance at the decoupling epoch $z\simeq1100$ but also the late integrated Sachs-Wolfe (ISW) effect. However, the late ISW effect cannot be at present accurately measured and we only focus on the smooth dark energy in the study, so the distance information given by the CMB distance priors is sufficient for the joint geometric constraint on dark energy. Thus, it is not necessary to use the full data of the CMB anisotropies. For an economical consideration, we decide to use the substracted information from CMB, i.e., the Planck distance priors.

\emph{The BAO data}: We use the recent BAO measurements from the Main Galaxy Sample of Data Release 7 of Sloan Digital Sky Survey (SDSS-MGS) at $z_{\rm eff} = 0.15$ \cite{Ross:2014qpa}, the six-degree-field galaxy survey (6dFGS)  measurement at $z_{\rm eff} = 0.106$ \cite{Beutler:2011hx}, and the CMASS and LOWZ samples from the Data Release 11 of the Baryon Oscillation Spectroscopic Survey (BOSS) at $z_{\rm eff} = 0.57$ and $z_{\rm eff} = 0.32$ \cite{Anderson:2013zyy}. This combination of BAO data has been used widely and proven to be in good agreement with the Planck CMB data.

\emph{The $H_{0}$ measurement}: We use the measurement of the Hubble constant in the light of the cosmic distance ladder, given by Efstathiou \cite{Efstathiou:2013via}, $H_{0}=70.6\pm3.3\rm \,km\, s^{-1} \rm Mpc^{-1}$, which is a re-analysis of the Cepheid data of Riess et al \cite{Riess:2011yx}.

See also Refs.~\cite{Wang:2013zca,Cui:2015oda,He:2016rvp,Zhang:2017rbg,Zhang:2017epd,Cui:2017idf,Hu:2015bpa,Wang:2016iij} for the use of the above observational data sets.

\section{Results and discussion}

\begin{table*}\small
\setlength\tabcolsep{5pt}
\caption{Fitting results of the parameters with best-fit values as well as 1$\sigma$ errors in all the I$\Lambda$CDM models.}
\label{table1}
\renewcommand{\arraystretch}{1.5}\centering
\begin{tabular}{ccccccccccc}
\\
\hline\hline
Parameter  & I$\Lambda$CDM1 & I$\Lambda$CDM2 & I$\Lambda$CDM3 & I$\Lambda$CDM4 & I$\Lambda$CDM5& I$\Lambda$CDM6\\ \hline

$\Omega_{\rm m}$      & $0.3152^{+0.0085}_{-0.0091}$
                   & $0.3150^{+0.0092}_{-0.0089}$
                   & $0.3140^{+0.0089}_{-0.0060}$
                   & $0.3143^{+0.0088}_{-0.0092}$
                   & $0.3164^{+0.0074}_{-0.0103}$
                   & $0.3137^{+0.0100}_{-0.0080}$
                   \\

$b_0$      & $-0.3295^{+0.3518}_{-0.2599}$
                   & $-0.4148^{+0.4517}_{-0.3603}$
                   & $-0.5296^{+0.3429}_{-0.3467}$
                   & $-0.2922^{+0.2725}_{-0.1957}$
                   & $-0.3115^{+0.3036}_{-0.3980}$
                   & $-0.4012^{+0.3944}_{-0.2795}$
                   \\

$b_e$                & $1.4079^{+1.0358}_{-1.4708}$
                   & $1.8372^{+1.6063}_{-1.9337}$
                   & $0.8647^{+0.5487}_{-0.5485}$
                   & $0.9848^{+0.6698}_{-0.9005}$
                   & $1.1329^{+1.5331}_{-1.0766}$
                   & $1.4669^{+1.0740}_{-1.4104}$
                   \\

$h$            & $0.6798^{+0.0107}_{-0.0119}$
                   & $0.6788^{+0.0104}_{-0.0110}$
                   & $0.7016^{+0.0224}_{-0.0235}$
                   & $0.6843^{+0.0123}_{-0.0150}$
                   & $0.6786^{+0.0136}_{-0.0098}$
                    & $0.6822^{+0.0097}_{-0.0130}$
                    \\

\hline
$\chi^2_{\rm min}$              & $698.35$
                   & $698.31$
                   & $697.35$
                   & $698.09$
                   & $698.17$
                   & $698.19$
                   \\
\hline\hline
\end{tabular}
\end{table*}

\begin{figure*}[tbp]
\centering % \begin{center}/\end{center} takes some additional vertical space
\includegraphics[width=13cm]{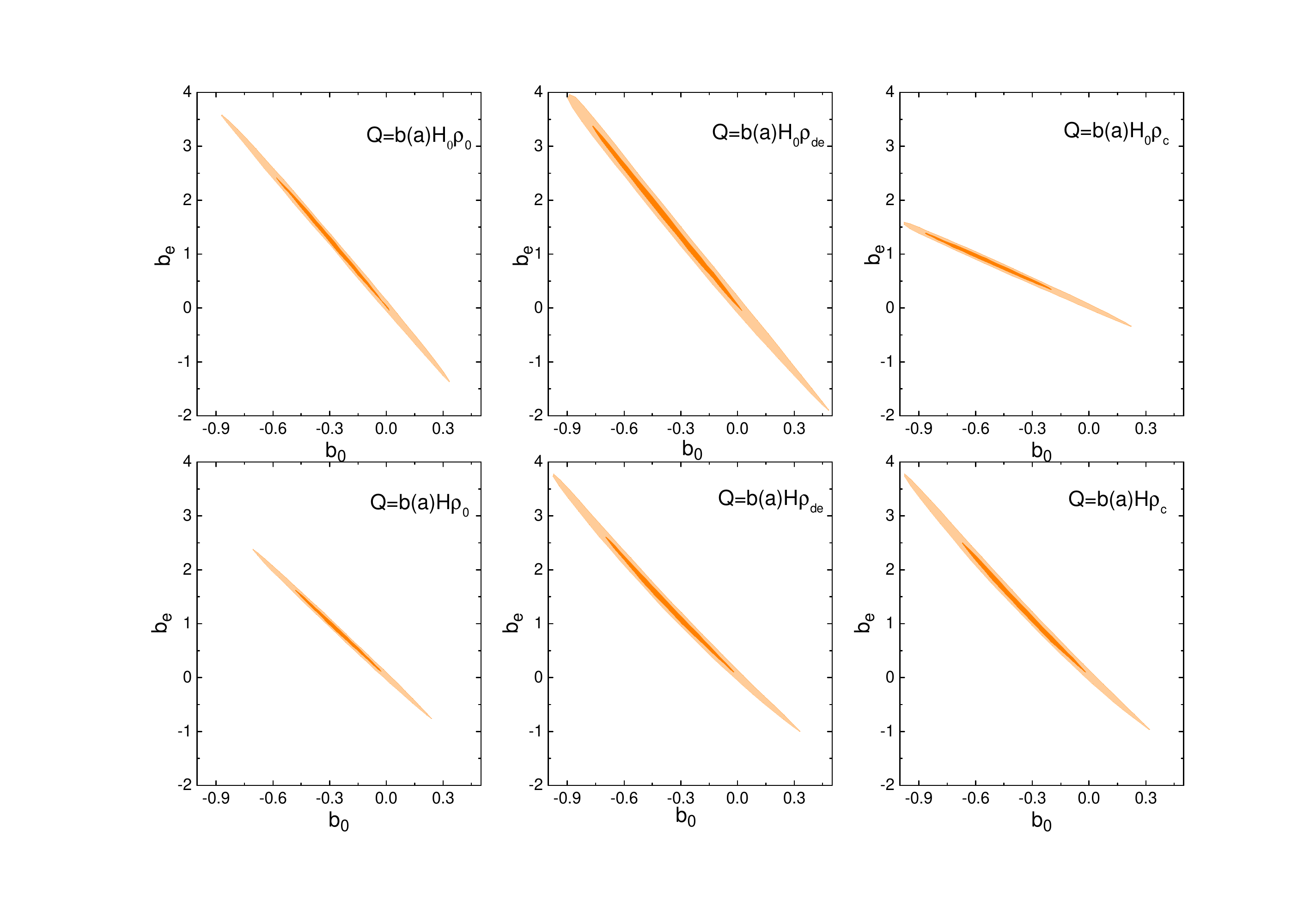}
\hfill
%\includegraphics[width=.45\textwidth,origin=c,angle=180]{img2.pdf}
% "\includegraphics" is very powerful; the graphicx package is already loaded
\caption{\label{fig1} The 1 $\sigma$ and 2 $\sigma$ confidence level contours in the $b_0$--$b_e$ plane, for the six I$\Lambda$CDM models, by using CMB+BAO+SN+$H_{0}$.}
%Constraints (68.3 and 95.4\% CL) in the $b_0$--$b_e$ plane for I$\Lambda$CDM1, I$\Lambda$CDM2, I$\Lambda$CDM3, I$\Lambda$CDM4, I$\Lambda$CDM5 and I$\Lambda$CDM6 models by using different data combinations..
%Constraints (68.3 and 95.4$\%$)results in the $n_t$--$r_{0.002}$ plane for the $\Lambda$CDM+$r$+$\nu_s$+$n_t$ model by using different data combinations. The black dashed line is plotted for the function $r_{0.002}=0.2(\frac{k=0.002}{k=0.01})^{n_t}$. The black dotted line denotes the consistency relation $r_{0.002}=-8n_t$.}
\end{figure*}

\begin{figure*}[tbp]
\centering % \begin{center}/\end{center} takes some additional vertical space
\includegraphics[width=13cm]{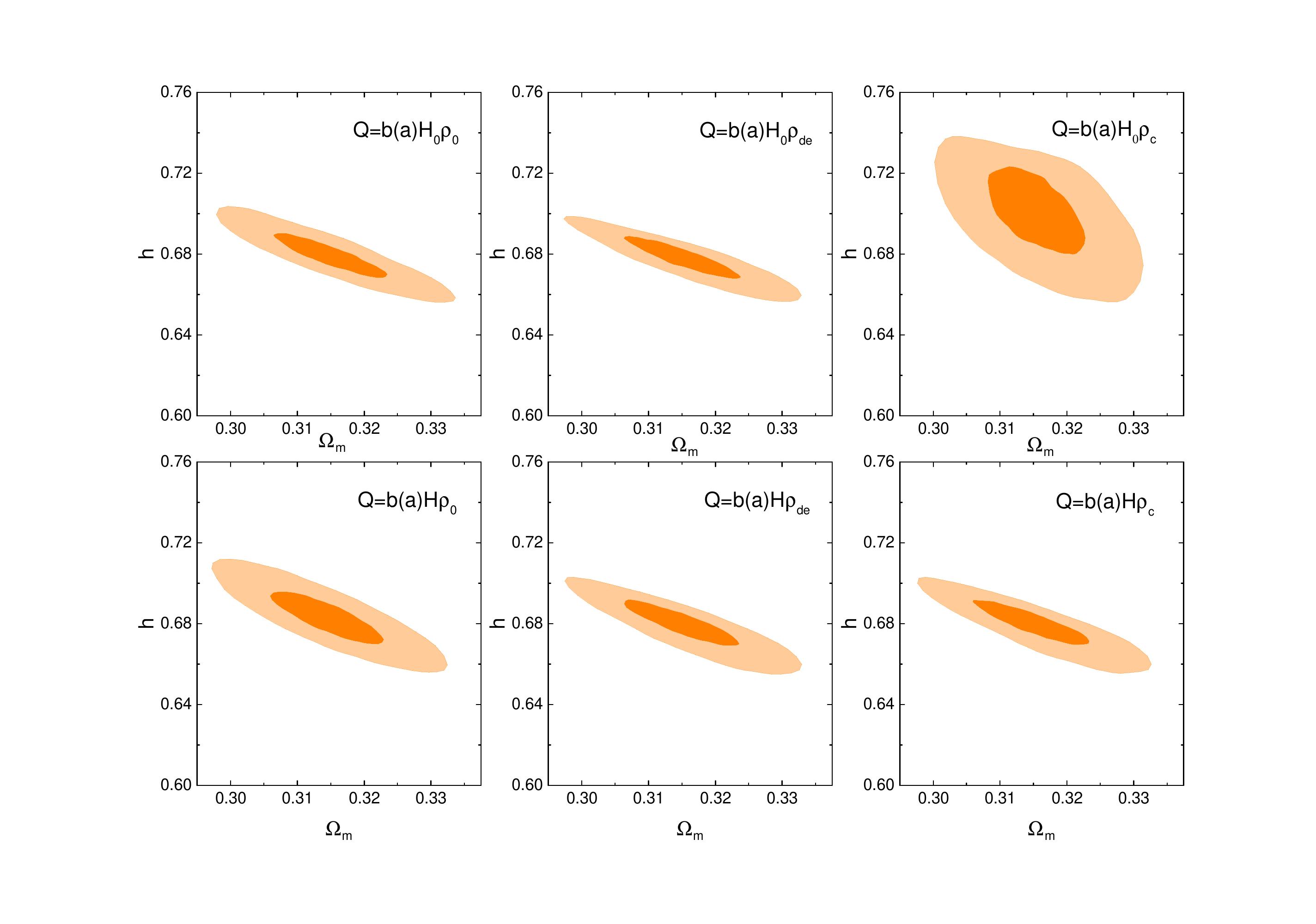}
\hfill
%\includegraphics[width=.45\textwidth,origin=c,angle=180]{img2.pdf}
% "\includegraphics" is very powerful; the graphicx package is already loaded
\caption{\label{fig2} The 1$\sigma$ and 2$\sigma$ confidence level contours in the $\Omega_m$--$h$ plane, for the six I$\Lambda$CDM models, by using CMB+BAO+SN+$H_{0}$.}
\end{figure*}

\begin{figure*}[tbp]
\centering % \begin{center}/\end{center} takes some additional vertical space
\includegraphics[width=13cm]{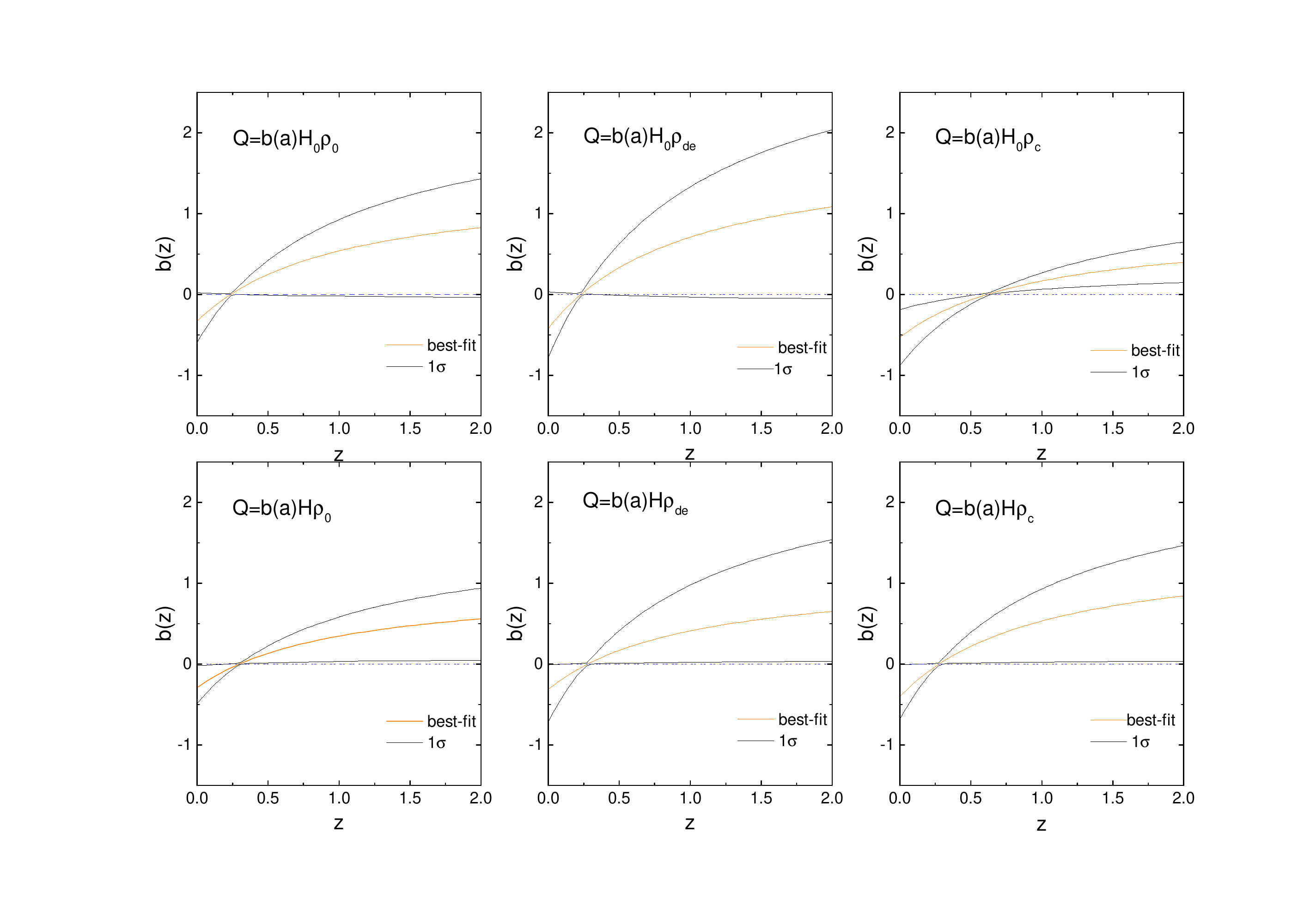}
\hfill
%\includegraphics[width=.45\textwidth,origin=c,angle=180]{img2.pdf}
% "\includegraphics" is very powerful; the graphicx package is already loaded
\caption{\label{fig3} The reconstructed evolutionary histories for $b(z)$ in the six I$\Lambda$CDM models. The black solid line represents the 1$\sigma$ uncertainties. The blue dashed line in each plot denotes the noniteracting line.}
\end{figure*}

We place constraints on the six I$\Lambda$CDM models by performing an MCMC likelihood analysis with the current joint CMB+BAO+SN+$H_{0}$ observations. We let eight chains run independently from each other so as to check the convergence and then obtain the fitting results of the best-fit, 1$\sigma$, and 2$\sigma$ values (and likelihood distributions) for the model parameters, as well as the corresponding $\chi^2_{\rm min}$ value for each model. The fit results are shown in Table~\ref{table1}. Here we also explicitly show these results:

\begin{itemize}

\item For the I$\Lambda$CDM1 model, we have $\Omega_{\rm m}=0.3152^{+0.0085}_{-0.0091}$, $b_{0}=-0.3295^{+0.3518}_{-0.2599}$, $b_e=1.4079^{+1.0358}_{-1.4708}$, and $h=0.6798^{+0.0107}_{-0.0119}$, with $\chi^2_{\rm min}=698.35$.
\item For the I$\Lambda$CDM2 model, we have $\Omega_{\rm m}=0.3150^{+0.0092}_{-0.0089}$, $b_{0}=-0.4148^{+0.4517}_{-0.3603}$, $b_e=1.8372^{+1.6063}_{-1.9337}$, and $h=0.6788^{+0.0104}_{-0.0110}$, with $\chi^2_{\rm min}=698.31$.
\item For the I$\Lambda$CDM3 model, we have $\Omega_{\rm m}=0.3140^{+0.0089}_{-0.0060}$, $b_{0}=-0.5296^{+0.3429}_{-0.3467}$, $b_e=0.8647^{+0.5487}_{-0.5485}$, and $h=0.7016^{+0.0224}_{-0.0235}$, with $\chi^2_{\rm min}=697.35$.
\item For the I$\Lambda$CDM4 model, we have $\Omega_{\rm m}=0.3143^{+0.0088}_{-0.0092}$, $b_{0}=-0.2922^{+0.2725}_{-0.1957}$, $b_e=0.9848^{+0.6698}_{-0.9005}$, and $h=0.6843^{+0.0123}_{-0.0150}$, with $\chi^2_{\rm min}=698.09$.
\item For the I$\Lambda$CDM5 model, we have $\Omega_{\rm m}=0.3164^{+0.0074}_{-0.0103}$, $b_{0}=-0.3115^{+0.3036}_{-0.3980}$, $b_e=1.1329^{+1.5331}_{-1.0766}$, and $h=0.6786^{+0.0136}_{-0.0098}$, with $\chi^2_{\rm min}=698.17$.
\item For the I$\Lambda$CDM6 model, we have $\Omega_{\rm m}=0.3137^{+0.0100}_{-0.0080}$, $b_{0}=-0.4012^{+0.3944}_{-0.2795}$, $b_e=1.4669^{+1.0740}_{-1.4104}$, and $h=0.6822^{+0.0097}_{-0.0130}$, with $\chi^2_{\rm min}=698.19$.

\end{itemize}

Figure~\ref{fig1} shows the joint constraints on the six I$\Lambda$CDM models in the $b_0$--$b_e$ plane. We can clearly see that $b_0<0$ and $b_e>0$ at about the 1$\sigma$ level for all the six models. The figure also explicitly shows that, for all the six models, $b_0$ and $b_e$ are in extremely strong anti-correlation. Actually, in the previous study \cite{Li:2011ga}, the anti-correlation between $b_0$ and $b_e$ was revealed. But the present work shows that the degree of the anti-correlation in the current case is much higher than the previous one. From Fig.~\ref{fig1}, we also find that, for all the models, it appears that $b_0$ and $b_e$ are in some linear relationship. We thus try to find out these linear relationships for them. Based on the likelihood contours, we can fit the equations of straight lines through the least square method using \texttt{Matlab}. By the calculation, we obtain the results:

\begin{itemize}
\item $b_e=-4.105b_0+0.038$ (I$\Lambda$CDM1),
\item $b_e=-4.249b_0+0.093$ (I$\Lambda$CDM2),
\item $b_e=-1.580b_0+0.025$ (I$\Lambda$CDM3),
\item $b_e=-3.317b_0+0.031$ (I$\Lambda$CDM4),
\item $b_e=-3.656b_0+0.076$ (I$\Lambda$CDM5),
\item $b_e=-3.659b_0+0.075$ (I$\Lambda$CDM6).
\end{itemize}

In Fig.~\ref{fig2}, we show the posterior distribution contours (1$\sigma$ and 2$\sigma$) in the $\Omega_{\rm m}$--$h$ plane. We find that, for all the models, $\Omega_{\rm m}$ and $h$ are anti-correlated, as the usual case in the base $\Lambda$CDM model.

We also find that the values of $\chi_{\rm min}^2$ for all the six models are almost equal (the difference can be ignored), indicating that the current observations equally favor these I$\Lambda$CDM models.

From Fig.~\ref{fig1}, we can also see that, in the I$\Lambda$CDM3 and I$\Lambda$CDM4 models, the constraints on $b_e$ are slightly tighter (see also Table.~\ref{table1}). In particular, we find that among these six models the I$\Lambda$CDM3 model is somewhat special. From the linear relationships between $b_0$ and $b_e$ obtained above, we find that, for the I$\Lambda$CDM3 model, the absolute value of the proportionality coefficient is the smallest (the proportionality coefficient is around $-1.5$ for the I$\Lambda$CDM3 model, and is around $-3.5\sim -4$ for other models). Thus, it can be clearly seen in Fig.~\ref{fig1} that the slope of the I$\Lambda$CDM3 model is evidently different. In Fig.~\ref{fig2}, we see that, in the I$\Lambda$CDM3 model, the range of $H_0$ is also evidently large.

In Fig.~\ref{fig3}, we reconstruct the evolution of $b(z)$ according to the fit results, in which the best fit and the 1$\sigma$ uncertainty are shown. Since $b_0<0$ and $b_e>0$ at around the 1$\sigma$ level are favored by the current data, for all the models, we show in Fig.~\ref{fig3} that $b(z)>0$ at the early times and $b(z)<0$ at the late times, and during the cosmological evolution the coupling $b(z)$ crosses the noninteracting line ($b=0$), at about the 1$\sigma$ level. For the I$\Lambda$CDM3 model, the crossing is favored at more than 1$\sigma$ level. The crossing happens at the redshift around 0.2--0.3 for almost all the models, except for the I$\Lambda$CDM3 model, for which it happens at $z\sim 0.6$. The investigations in the present work confirm the conclusion in Ref.~\cite{Li:2011ga} that the energy transfer is from DM to DE when DM dominates the universe and the energy transfer is from DE to DM when DE dominates the universe. We investigate more interaction forms and find that their results are in good accordance.

\section{Conclusion}

Using the observational data to test the phenomenological interacting dark energy scenario is a significant mission in cosmology today. So far, this issue has been addressed widely. In this paper, we wish to explore the issue whether the coupling between DE and DM would change sign during the cosmic evolution by using the current joint geometric measurement data.

In the previous work \cite{Li:2011ga}, the authors put forward a running coupling scenario in which $b(a)=b_0a+b_e(1-a)$. The investigation in Ref.~\cite{Li:2011ga} indicated that a time-varying vacuum scenario is favored, in which the coupling $b(z)$ crosses the non-interacting line during the cosmic evolution. In this paper, in order to check the conclusion and complete the study in the current situation that we have more accurate observational data, we have investigated six I$\Lambda$CDM models, i.e., the I$\Lambda$CDM1 model with $Q=b(a)H_0\rho_0$, the I$\Lambda$CDM2 model with $Q=b(a)H_0\rho_{\rm de}$, the I$\Lambda$CDM3 model with $Q=b(a)H_0\rho_{\rm c}$, the I$\Lambda$CDM4 model with $Q=b(a)H\rho_{0}$, the I$\Lambda$CDM5 model with $Q=b(a)H\rho_{\rm de}$ and the I$\Lambda$CDM6 model with $Q=b(a)H\rho_{\rm c}$. We place stringent constraints on the models by using the current combined data, including the JLA SN data, the Planck CMB distance priors data, the BAO data, and the $H_{0}$ direct measurement.

The results of observational constraints for these models are given in Table~\ref{table1} and Figs.~\ref{fig1} and \ref{fig2}. With the purpose of visually displaying the coupling $b(z)$ crossing the non-interacting line ($b=0$), we reconstructed the evolution of the coupling $b(z)$ with $1\sigma$ uncertainties in Fig.~\ref{fig3}.

Our results show that, for all the I$\Lambda$CDM models, we have $b_0<0$ and $b_e>0$ at around the 1$\sigma$ level, and $b_0$ and $b_e$ are in extremely strong anti-correlation. Since $b_0$ and $b_e$ seem to be in some linear relationship, we have tried to find out these linear relationships for them. Though the I$\Lambda$CDM3 model appears to be somewhat special among these models, the main conclusions for all the models are in good accordance. The reconstructed evolutions of $b(z)$ explicitly show that the coupling changes sign during the evolution, i.e., $b(z)>0$ at the early times and $b(z)<0$ at the late times, indicating that the energy transfer is from DM to DE when DM dominates the universe and the energy transfer is from DE to DM when DE dominates the universe. Our results confirm the conclusion of Ref.~\cite{Li:2011ga}. Since in this study we have explored more models and used the latest observations, the more convincing and conclusive evidence is provided.

\begin{acknowledgments}
This work was supported by the National Natural Science Foundation of China (Grants No.~11522540 and No.~11690021), the National Program for Support of Top-Notch Young Professionals, and the Provincial Department of Education of Liaoning (Grant No.~L2012087).
\end{acknowledgments}

\end{document}